\documentclass[11pt,a4paper]{article}

\usepackage{jheppub}
\usepackage[utf8]{inputenc}
\usepackage[english]{babel}
\usepackage{braket}
\usepackage{verbatim}
\usepackage{empheq}

\newcommand{\bm}{\boldsymbol}
\newcommand{\st}{{\scriptscriptstyle T}}
\def\nn{\nonumber}
\def\cd{{\cdot}}
\DeclareMathOperator{\tr}{Tr}

\usepackage[normalem]{ulem}
\renewcommand\sout{\bgroup \color[rgb]{0.55,0.00,0.99} \ULdepth=-.5ex \ULset}

\allowdisplaybreaks

\def\be{\begin{equation}}
\def\ee{\end{equation}}
\def\bea{\begin{eqnarray}}
\def\eea{\end{eqnarray}}

\begin{document}

\preprint{NIKHEF 2018-025}

\title{Directed flow from C-odd gluon correlations at small $\bm{x}$}

\author[a]{Dani\"el Boer,}
\author[b,c]{Tom van Daal,}
\author[b,c]{Piet J. Mulders,}
\author[b,c]{Elena Petreska}

\affiliation[a]{Van Swinderen Institute for Particle Physics and Gravity, University of Groningen, Nijenborgh 4, NL-9747 AG Groningen, The Netherlands}
\affiliation[b]{Department of Physics and Astronomy, VU University Amsterdam, De Boelelaan 1081, NL-1081 HV Amsterdam, The Netherlands}
\affiliation[c]{Nikhef, Science Park 105, NL-1098 XG Amsterdam, The Netherlands}

\emailAdd{d.boer@rug.nl}
\emailAdd{tvdaal@nikhef.nl}
\emailAdd{mulders@few.vu.nl}
\emailAdd{petreska@nikhef.nl}

\abstract{It is shown that odd harmonic azimuthal correlations, including the directed flow $v_1$, in forward two-particle production in peripheral proton-nucleus ($pA$) collisions can arise simply from the radial nuclear profile of a large nucleus. This requires consideration of the C-odd part of the gluonic generalized transverse momentum dependent (GTMD) correlator of nucleons in the nucleus. The gluonic GTMD correlator is the Fourier transform of an off-forward hadronic matrix element containing gluonic field strength tensors that are connected by gauge links. It is parametrized in terms of various gluon GTMD distribution functions (GTMDs). We show (in a gauge invariant way) that for the relevant dipole-type gauge link structure in the small-$x$ limit the GTMD correlator reduces to a generalized Wilson loop correlator. The Wilson loop correlator is parametrized in terms of a single function, implying that in the region of small $x$ there is only one independent dipole-type GTMD, which can have a C-odd part. We show that the odderon Wigner distribution, which is related to this C-odd dipole GTMD by a Fourier transform, generates odd harmonics in the two-particle azimuthal correlations in peripheral $pA$ collisions. We calculate the first odd harmonic $v_1$ for forward production within the color glass condensate framework in the limit of a large number of colors. We find that nonzero odd harmonics are present without breaking the rotational symmetry of the nucleus, arising just from its inhomogeneity in the radial direction. Using a CGC model with a cubic action, we illustrate that percent level $v_1$ can arise from this C-odd mechanism. In contrast, we show that only even harmonics arise in diffractive dijet production in ultra-peripheral $pA$ collisions where this gluon dipole GTMD also appears.}


\arxivnumber{1805.05219}

\maketitle

\flushbottom

\section{Introduction}
Generalized transverse momentum dependent distributions (GTMDs) of partons inside hadrons are off-forward matrix elements that combine information on transverse momentum dependent distributions (TMDs) and generalized parton distributions (GPDs)~\cite{Ji:2003ak,Belitsky:2003nz}. The (forward) TMDs depend on the longitudinal momentum fraction $x$ as well as on the partonic transverse momentum $\bm{k}^2$. The GPDs, on the other hand, involve off-forward states (i.e.\ $\Delta \equiv p'-p \ne 0$) and depend on $x$, the skewness parameter $\xi$, and the transverse momentum transfer $\bm{\Delta}^2$. The GTMDs appear in the parametrization of the GTMD correlator. The parametrization for the quark case is given in~\cite{Meissner:2008ay,Meissner:2009ww} and~\cite{Lorce:2013pza} covers also the gluon case. The Fourier transform of the GTMDs from $\bm{\Delta}$ to the impact parameter $\bm{b}$, gives the Wigner distributions which provide information on both the spatial and momentum structure of hadrons. They are the quantum mechanical analogues of the classical phase-space distributions of the parton-hadron system. We will discuss gluon (rather than quark) GTMDs and Wigner distributions as we will focus on the small-$x$ kinematic region where gluons dominate the hadron structure.

The gluon-gluon (G)TMD correlator is the Fourier transform of a bilocal hadronic matrix element of two field strength tensors. The nonlocality includes transverse directions and is bridged by gauge links that ensure color gauge invariance~\cite{Belitsky:2002sm,Boer:2003cm}. The integration paths of the gauge links are not unique, but in fact depend on the process~\cite{Collins:1983pk,Boer:1999si,Collins:2002kn}. The most important gauge link structures involve the so-called future- and past-pointing staple-like gauge links $U^{[+]}$ and $U^{[-]}$ respectively. The well-known unintegrated dipole gluon distribution~\cite{Kharzeev:2003wz} features both of these links~\cite{Dominguez:2010xd,Dominguez:2011wm}. For the dipole-type gauge link structure, the gluon-gluon TMD correlator simplifies in the small-$x$ limit to the Fourier transform of a hadronic matrix element containing a single Wilson loop. It was recently shown in~\cite{Boer:2015pni,Boer:2016xqr} that this Wilson loop correlator can be used to greatly reduce the number of independent TMDs in the small-$x$ limit. We will show that a similar thing can be done for the GTMD case as well. With respect to~\cite{Lorce:2013pza}, we provide an alternative parametrization of the gluon-gluon GTMD correlator (for unpolarized hadrons) in terms gluon GTMDs. We show that in the limit of small $x$ and $\xi$ we are left with only one independent dipole-type gluon GTMD. 

In the second part of the paper, we will consider the dipole-type gluon GTMDs and Wigner distributions within the color glass condensate (CGC) framework~\cite{Gelis:2010nm}. In particular, we will study those distributions as sources for angular correlations that have been observed experimentally in proton-proton ($pp$) and proton-nucleus ($pA$) collisions at the LHC~\cite{Abelev:2012ola,ABELEV:2013wsa,Abelev:2014mda,Adam:2015bka,Acharya:2018dxy,Aad:2012gla,Aad:2013fja,Aad:2014lta,Aad:2015gqa,Aaboud:2016yar,Aaboud:2017acw,Aaboud:2017blb,Khachatryan:2010gv,Khachatryan:2015waa,CMS:2012qk,Chatrchyan:2013nka,Khachatryan:2015oea,Khachatryan:2015lva,Khachatryan:2016txc,Sirunyan:2017igb,Sirunyan:2017uyl,Aaij:2015qcq} and RHIC~\cite{Adare:2013piz,Adare:2014keg,Belmont:2017lse,Adamczyk:2014fcx,Adamczyk:2015xjc}. The source of large azimuthal asymmetries in hadron production in $pp$ and $pA$ collisions has been widely addressed in the CGC framework and especially the initial-state origin of azimuthal harmonics (as opposed to final-state interactions and hydrodynamic flow) has been studied~\cite{Kovchegov:1999ep,Teaney:2002kn,Kovchegov:2002nf,Dumitru:2008wn,Gavin:2008ev,Avsar:2010rf,Dumitru:2010iy,Kovner:2010xk,Kovner:2011pe,Levin:2011fb,Iancu:2011ns,Schenke:2012wb,Schenke:2012fw,Dusling:2012iga,Dusling:2013qoz,Dusling:2015rja,Kovchegov:2013ewa,Kovner:2012jm,Kovner:2014qea,Dumitru:2014yza,Kovner:2015rna,Schenke:2015aqa,Lappi:2015vha,Lappi:2015vta,Altinoluk:2015dpi,Rezaeian:2016szi,Schenke:2016ksl,Schenke:2016lrs,Dusling:2017dqg,Dusling:2017aot,Fukushima:2017mko,Altinoluk:2018hcu}. Here, we concentrate on generating nonzero odd harmonics entirely from initial-state effects which is less understood compared to generating even harmonics. The natural way of obtaining odd azimuthal asymmetries for quarks scattering off the CGC field comes from the C-odd imaginary part of the fundamental dipole scattering amplitude, called the dipole odderon. The cross section for hadron production from an incoming quark is proportional to the Fourier transform of the expectation value of the dipole operator $S = \tr [U(\bm{x}) \,U^\dagger(\bm{y})] /N_c$, where $U(\bm{x})$ is a fundamental Wilson line. The real part of $S$ produces even harmonics, while the imaginary part (the odderon) generates odd harmonics. Odd harmonics for quark scattering, from initial-state effects, have been studied in~\cite{Dumitru:2014dra,Dumitru:2014vka,Lappi:2015vta,Lappi:2015vha}. In those works, the odd angular coefficients vanish once the contribution from antiquarks is taken into account. For the case of gluon scattering, odd harmonics in double-inclusive gluon production from initial-state effects were recently introduced in~\cite{Kovner:2016jfp,Kovchegov:2018jun}.

In this paper we study odd azimuthal correlations between two hadrons produced in $pA$ collisions and connect them to the odderon Wigner distribution (the impact parameter dependent odderon). We mostly restrict the analysis to forward rapidities and consider dihadron production from two incoming quarks from the proton (the quarks are in the valence region) that interact with a large nucleus. The scatterings of the quarks are taken as independent and only the leading contribution in the number of colors (disconnected diagrams) is considered. Our analysis follows the work of~\cite{Hagiwara:2017ofm}, extending it to include the odderon Wigner distribution and showing the appearance of nonzero odd harmonics. We show that the odderon Wigner distribution at small $x$ is related to the $v_1$ flow coefficient (that characterizes the so-called directed flow) in dihadron production in $pA$ collisions. Compared to previous literature, the source of odd azimuthal asymmetries in our model is the relative orientation between the impact parameter and the transverse momentum of the produced particle for a nucleus that is inhomogeneous in the transverse plane. The same mechanism was discussed in~\cite{Iancu:2017fzn} for the case of even harmonics and for single-inclusive particle production. The connection between even azimuthal asymmetries and the orientation of the dipole with respect to its impact parameter was studied in~\cite{Kopeliovich:2007fv,Kopeliovich:2007sd,Kopeliovich:2008nx,Levin:2011fb}. Although $v_1$ can arise from C-odd fluctuations in an anisotropic rotationally noninvariant target~\cite{Dumitru:2014dra}, we show that it can also arise from C-odd correlations without breaking of rotational symmetry. In the latter case it enters as a C-odd squared effect. We derive an explicit expression for the $v_1$ flow coefficient in the semi-classical limit of the CGC theory in the limit of a large number of colors ($N_c$) and observe a nonzero effect in the region of small transverse momenta with the same order of magnitude as is experimentally observed. We also argue that away from forward rapidities, in which case also antiquarks in the proton contribute, the same mechanism gives rise to odd azimuthal asymmetries.

For comparison, we also consider the production of two jets in diffractive DIS and in ultra-peripheral $pA$ collisions. The cross section for these processes involves the gluon dipole GTMD~\cite{Altinoluk:2018hcu,Hatta:2016dxp,Hagiwara:2017fye}, but unlike dihadron production in $pA$ collisions, we find that the odderon GTMD does not produce odd azimuthal asymmetries and only contributes to the even harmonics.

The paper is organized as follows. In section~\ref{sec:Parametrization} we parametrize the gluon-gluon GTMD correlator for an unpolarized hadron in terms of gluon GTMDs. Subsequently, in section~\ref{sec:WilsonLoop} we study the dipole-type GTMD correlator and find that for vanishing $x$ and $\xi$ there is only one independent dipole GTMD. In section~\ref{s:AzimuthalCorr.} we study odd azimuthal correlations in two-particle production in $pA$ collisions and we calculate $v_1$ within the CGC framework. Finally, we conclude in section~\ref{Conclusions}.

\section{Parametrization of the GTMD correlator} \label{sec:Parametrization}
\begin{figure}[!htb]
\begin{center}
    \begin{minipage}{8cm}
    \hspace{0.35cm}\includegraphics[width=\textwidth]{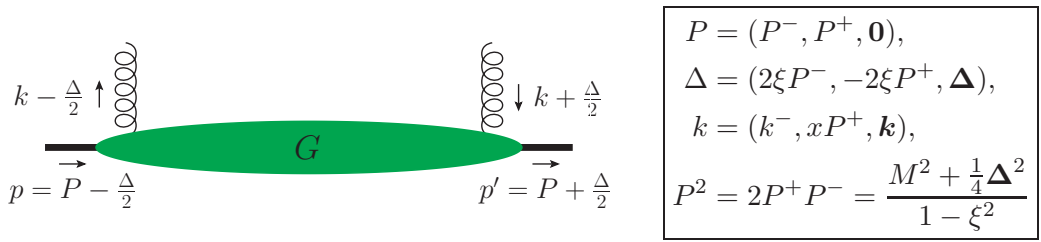}
    \end{minipage}
    \quad
    \begin{minipage}{6cm} \vspace{-0.4cm}
    \begin{empheq}[box=\fbox]{align*}
    P &= (P^-,P^+,\bm 0) , \\
    \Delta &= (2\xi P^-,-2\xi P^+,\bm{\Delta}) , \\
    k &= (k^-,xP^+,\bm k) , \\
    P^2 &= 2P^+P^- = \frac{M^2 + \tfrac{1}{4}\bm{\Delta}^2}{1-\xi^2}
    \end{empheq}
    \end{minipage}
\end{center}
\vspace{-0.4cm}
\caption{The kinematical set-up; the green `blob' represents the gluon-gluon GTMD correlator.}
\label{f:correlator}
\end{figure}

We parametrize the momenta in a `symmetric' way where the average hadron momentum is given by $P \equiv (p'+p)/2$, the momentum transfer by $\Delta \equiv p'-p$, and the average gluon momentum is denoted by $k$, see also figure~\ref{f:correlator}. Furthermore, in our applications the momenta $p$ and $p'$ can be taken to have large plus components $p^+ = p{\cdot}n$ and $p'^+ = p'{\cdot}n$. In figure~\ref{f:correlator} the momenta $P$, $\Delta$, and $k$ are given in terms of light-cone components.\footnote{Throughout the paper we make use of light-cone coordinates: we represent a four-vector $a$ as $(a^-, a^+, \bm{a})$, where $a^\pm \equiv (a^0 \pm a^3)/\sqrt{2}$ and $\bm{a} \equiv (a^1, a^2)$. We also define the four-vector $a_\st \equiv (0, 0, \bm{a})$, so that $a_\st^2 = -\bm{a}^2$.} We take hadrons of the same mass, $p^2 = p'^2 = M^2$, the light-cone fraction of $k$ is defined as $x \equiv k^+/P^+$, and the skewness parameter is given by $2\xi \equiv -\Delta^+/P^+$. 

The gluon-gluon GTMD correlator for an unpolarized hadron generalizes the TMD correlator~\cite{Bomhof:2006dp} and is defined as~\cite{Lorce:2013pza}
\begin{align}
    &\,G^{[U,U^\prime]\,\mu\nu;\rho\sigma}(x,\bm{k},\xi,\bm{\Delta};P,n) \nn \\
    \equiv& \,\frac{2}{P^+} \left. \int \frac{dz^- \,d^2\bm{z}}{(2\pi)^3} \;e^{ik\cdot z} \bra{p'} \tr \left( F^{\mu\nu} \!\left(-\tfrac{z}{2}\right) U_{\left[-\tfrac{z}{2},\tfrac{z}{2}\right]}^{\phantom{\prime}} \,F^{\rho\sigma} \!\left(\tfrac{z}{2}\right) U_{\left[\tfrac{z}{2},-\tfrac{z}{2}\right]}^\prime \right) \ket{p} \right|_{z^+ = 0} .
\end{align}
Similar to the TMD case, it was recently shown that a proper definition of the GTMD correlator includes additional dependence on soft-gluon radiation~\cite{Echevarria:2016mrc}, however as this will not play any role in this paper it is not discussed here explicitly. The two process-dependent gauge links $U$ and $U'$ are required for color gauge invariance. Later we will consider a specific gauge link structure that is built from the future- and past-pointing gauge links $U_{[a,b]}^{[+]}$ and $U_{[a,b]}^{[-]}$ respectively, defined as 
\begin{equation}
    U_{[a,b]}^{[\pm]} \equiv U_{[a^-,\pm \infty;\bm{a}]}^n \,U_{[\pm \infty;\bm{a},\bm{b}]}^T \,U_{[\pm \infty,b^-;\bm{b}]}^n ,
    \label{e:WLs}
\end{equation}
where the links along the minus (indicated by the lightlike vector $n$) and transverse directions are given by
\begin{align}
    U_{[a^-,\pm \infty;\bm{a}]}^n \equiv \mathcal{P} \exp\left[ -ig \int_{a^-}^{\pm \infty} d\eta^- A^+(\eta^+=0,\eta^-,\bm{\eta}=\bm{a}) \right] , \label{e:U^n} \\
    U_{[\pm \infty;\bm{a},\bm{b}]}^T \equiv \mathcal{P} \exp\left[ -ig \int_{\bm{a}}^{\bm{b}} d\bm{\eta} \cdot \bm{A}(\eta^+=0,\eta^-=\pm \infty,\bm{\eta}) \right] ,
\end{align}
and likewise for the third factor in eq.~\eqref{e:WLs}. 

Counting powers of the inverse hard scale relevant in the process, leads to the definition of the leading-power (usually referred to as leading-twist) correlator:
\begin{equation}
    G^{[U,U^\prime]\,ij}(x,\bm{k},\xi,\bm{\Delta}) \equiv G^{[U,U^\prime]\,+i;+j}(x,\bm{k},\xi,\bm{\Delta};P,n) ,
    \label{e:W_GTMD}
\end{equation}
where $i,j$ are transverse indices and for convenience we suppress the dependence on $P$ and $n$. The correlator in eq.~\eqref{e:W_GTMD} can be parametrized in terms of GTMDs. This has been done already in~\cite{Lorce:2013pza} based on the light-front formalism (in fact, up to all powers in the inverse hard scale and including vector polarization) and here we present an alternative, but equivalent, parametrization. It is convenient to parametrize correlators in terms of symmetric traceless tensors to ensure that the distribution functions are of definite rank~\cite{Boer:2016xqr,Signori:2016lvd,vanDaal:2016glj}. In the GTMD case, one should use symmetric traceless tensors in both $k_\st$ and $\Delta_\st$. Hermiticity and time reversal (unlike parity) do not affect the Lorentz structure of the parametrization; they rather impose constraints on the GTMDs~\cite{Lorce:2013pza}. A possible parametrization is given by:\footnote{In principle, one could also have a function that comes with the symmetric and traceless Lorentz structure $k_\st^{\{i} \Delta_\st^{j\}} + (\bm{k} \cd \bm{\Delta}) \,g_\st^{ij}$. However, this function would not be independent from the other ones; more specifically, it could be eliminated by suitable redefinitions of the functions $\mathcal{F}_2$ and $\mathcal{F}_3$.}
\begin{align}
    G^{[U,U^\prime]\,ij}(x,\bm{k},\xi,\bm{\Delta}) &= x \left( \delta_\st^{ij} \,\mathcal{F}_1 + \frac{k_\st^{ij}}{M^2} \,\mathcal{F}_2 + \frac{\Delta_\st^{ij}}{M^2} \,\mathcal{F}_3 + \frac{k_\st^{[i} \Delta_\st^{j]}}{M^2} \,\mathcal{F}_4 \right) ,
    \label{e:W_GTMD_par}
\end{align}
where $\delta_\st^{ij}$ is the Kronecker delta in transverse space, the rank-2 symmetric traceless tensors $k_\st^{ij}$ and $\Delta_\st^{ij}$ are defined as $a_\st^{ij} \equiv a_\st^i a_\st^j + \frac{1}{2} \bm{a}^2 g_\st^{ij}$, and the square brackets denote antisymmetrization of the indices. The functions $\mathcal{F}_i = \mathcal{F}_i^{[U,U^\prime]}(x,\xi,\bm{k}^2,\bm{\Delta}^2,\bm{k} \cd \bm{\Delta})$ are complex-valued GTMDs; they are related to the leading-twist TMDs~\cite{Mulders:2000sh} and GPDs~\cite{Meissner:2007rx} for unpolarized hadrons upon setting $\Delta=0$ and integrating over $\bm{k}$ respectively.

\section{Wilson loop correlators and dipole-type GTMDs at small $\bm{x}$} \label{sec:WilsonLoop}
In order to study the dipole-type GTMD correlator, we start with the correlator that has a link structure involving both a future- and a past-pointing Wilson line as given in eq.~\eqref{e:WLs}. This link structure appears in the dipole gluon distribution, which is one of the distributions that describes unpolarized gluons inside unpolarized hadrons~\cite{Kharzeev:2003wz,Dominguez:2011wm}. The correlator is written as
\begin{align}
    G^{[+,-]\,ij}(x,\bm{k},\xi,\bm \Delta) &= \frac{2}{P^+} \int \frac{dz^- d^2\bm{z}}{(2\pi)^3} \;e^{ik\cdot z} \nn \\
    &\quad\, \times \left. \bra{p'} \tr \left( F^{+j} \!\left(\tfrac{z}{2}\right) U_{\left[-\tfrac{z}{2},\tfrac{z}{2}\right]}^{[-]\dag} \,F^{+i} \!\left(-\tfrac{z}{2}\right) U_{\left[-\tfrac{z}{2},\tfrac{z}{2}\right]}^{[+]} \right) \ket{p} \vphantom{\int} \right|_{\rm LF} \nn \\
    &= \frac{2}{P^+} \int \frac{dz^- d^2\bm{z}}{(2\pi)^3} \;e^{ik\cdot z- i\Delta\cdot b} \nn \\
    &\quad\, \times \left. \bra{p'} \tr \left( F^{+j} \!\left(x\right) U_{\left[y,x\right]}^{[-]\dag} \,F^{+i} \!\left(y\right) U_{\left[y,x\right]}^{[+]} \right) \ket{p} \vphantom{\int} \right|_{\rm LF} ,
\end{align}
where $x \equiv b + z/2$ and $y \equiv b - z/2$, and `LF' indicates the light front ($z^+ = x^+ = y^+ = 0$). Giving the operator combination an overall shift $b$ corresponds to a phase in the off-forward case. Using that the limit $\Delta \to 0$ of $\braket{p'|p} = (2\pi)^3 \,2p^+ \delta(\Delta^+) \,\delta^{(2)}(\bm{\Delta})$ is given by $\braket{P|P} = 2P^+ \int db^- d^2\bm{b}$, we can write
\begin{align}
    G^{[+,-]\,ij}(x,\bm{k},\xi,\bm\Delta) &= 4 \int \frac{d^3x \,d^3y}{(2\pi)^3} \;e^{ik\cdot (x - y) - i\Delta \cdot \frac{x+y}{2}} \nn \\
    &\quad\, \times \frac{\left. \bra{p'} \tr \left( F^{+j}(x) \,U_{[x,y]}^{[-]} \,F^{+i}(y) \,U_{[y,x]}^{[+]} \right) \ket{p} \right|_{\rm LF}}{\braket{P|P}} .
\end{align}
In the rest of this paper we will consider this GTMD correlator in the small-$x$ kinematic region where the gluonic content of hadrons is particularly relevant, i.e.\ the region where $x \sim \xi$ is smaller than the square of the hadronic scale divided by the relevant hard scale, which includes both the DGLAP region (for which $|\xi| < x$) and the ERBL region (for which $x < |\xi|$)~\cite{Diehl:2003ny}. Let us consider the limit
\begin{equation}
    G^{[+,-]\,ij}(\bm{k},\bm{\Delta}) \equiv \lim_{x,\xi \to 0} G^{[+,-]\,ij}(x,\bm{k},\xi,\bm{\Delta}) ,
    \label{e:correlator_atsmallx}
\end{equation}
in which case the $x^-$ and $y^-$ integrations can be performed. The correlator depends on the virtuality $k^2 = -\bm{k}^2$, the momentum transfer $t \equiv \Delta^2 = -\bm{\Delta}^2$, and the azimuthal angle between these spacelike transverse vectors. Similar to the TMD case in~\cite{Boer:2016xqr}, the minus integrations are used to express the result in terms of gluonic pole operators,
\begin{equation}
    G_\st^\alpha(x) \equiv \frac{1}{2} \int_{-\infty}^\infty d\eta^- \,U_{[x^-,\eta^-;\bm{x}]}^n \,F^{+\alpha}(x^+,\eta^-,\bm{x}) \,U_{[\eta^-,x^-;\bm{x}]}^n ,
\end{equation} 
which upon using the relation~\cite{Buffing:2013kca}
\begin{equation}
    \left[ i \partial_{\bm{x}}^\alpha, U_{[a,x]}^{[\pm]} \right] = \pm \,g \,U_{[a,x]}^{[\pm]} \,G_\st^\alpha(x) ,
\end{equation}
leads for the dipole-type GTMD correlator to a correlator of a Wilson loop:
\begin{align}
    G^{[+,-]\,ij}(\bm{k},\bm{\Delta}) &= 16 \int \frac{d^2\bm{x} \,d^2\bm{y}}{(2\pi)^3} \;e^{- i\bm{k} \cd (\bm{x} - \bm{y}) + i\bm{\Delta} \cd \frac{\bm{x}+\bm{y}}{2}} \nn \\
    &\quad\, \times \frac{\left. \bra{p'}\tr \left( \,G_\st^j(x) \,U_{[x,y]}^{[-]} \,G_\st^i(y) \,U_{[y,x]}^{[+]} \right) \ket{p} \right|_{\rm LF}}{\braket{P|P}} \nn \\
    &= \frac{4}{g^2} \int \frac{d^2\bm{x} \,d^2\bm{y}}{(2\pi)^3} \;e^{- i\bm{k} \cd (\bm{x} - \bm{y}) + i\bm{\Delta} \cd \frac{\bm{x}+\bm{y}}{2}} \,\frac{\left. \bra{p'} \,\partial_{\bm{x}}^j \partial_{\bm{y}}^i \tr \left( U^{[\Box]} \right) \ket{p} \right|_{\rm LF}}{\braket{P|P}} ,
    \label{e:W_GTMD_dipole}
\end{align}
where now the off-forwardness is only transverse ($\bm \Delta$) and where $U^{[\Box]} \equiv U_{[y,x]}^{[+]} \,U_{[x,y]}^{[-]}$ is a rectangular Wilson loop with transverse orientation $\bm{r} \equiv \bm{x} - \bm{y}$, stretched to infinity in the longitudinal direction. We can apply partial integration twice and write eq.~\eqref{e:W_GTMD_dipole} as\footnote{Eq.~\eqref{e:G_GTMD_S} is consistent with the result in~\cite{Hatta:2016dxp} where only the term with $\delta_\st^{ij}$ was considered.}
\begin{equation}
    G^{[+,-]\,ij}(\bm{k},\bm{\Delta}) = \frac{2N_c}{\alpha_s} \left[ \frac{1}{2} \left( \bm{k}^2 - \frac{\bm{\Delta}^2}{4} \right) \delta_\st^{ij} + k_\st^{ij} - \frac{\Delta_\st^{ij}}{4} - \frac{k_\st^{[i} \Delta_\st^{j]}}{2} \right] G^{[\Box]}(\bm{k},\bm{\Delta}) ,
    \label{e:G_GTMD_S}
\end{equation}
where
\begin{equation}
    G^{[\Box]}(\bm{k},\bm{\Delta}) \equiv \int \frac{d^2\bm{x} \,d^2\bm{y}}{(2\pi)^4} \;e^{- i\bm{k} \cd (\bm{x} - \bm{y}) + i\bm{\Delta} \cd \frac{\bm{x}+\bm{y}}{2}} \,\frac{\left. \bra{p'} S^{[\Box]}(\bm{x},\bm{y}) \ket{p} \right|_{\rm LF}}{\braket{P|P}} ,
    \label{e:G_at_small_x}
\end{equation}
is the Fourier transform of the Wilson loop operator
\begin{equation}
    S^{[\Box]}(\bm{x},\bm{y}) \equiv \frac{1}{N_c} \tr \left( U^{[\Box]} \right) .
    \label{e:dipole_operator}
\end{equation}
The off-forward correlator $G$ in eq.~\eqref{e:G_at_small_x} is the generalized Wilson loop correlator of which the forward limit is the correlator discussed in~\cite{Boer:2016xqr}. The Wilson loop operator in eq.~\eqref{e:dipole_operator} is the color gauge invariant version of the dipole operator $\tr [U(\bm{x}) \,U^\dagger(\bm{y})] /N_c$, which defines the $S$-matrix for a quark-antiquark pair scattering on a nucleus with the quark at position $\bm{x}$ and the antiquark at position $\bm{y}$.

Similar to the TMD case in~\cite{Boer:2016xqr}, the Wilson loop correlator can be parametrized in terms of GTMDs. We choose the following parametrization:
\begin{equation}
    G^{[\Box]}(\bm{k},\bm{\Delta}) = \frac{\alpha_s}{2N_c M^2} \;\mathcal{E}(\bm{k}^2,\bm{\Delta}^2,\bm{k} \cd \bm{\Delta}) ,
    \label{e:par_of_G}
\end{equation}
where $\mathcal{E}$ is a Wilson loop GTMD. Hence, at vanishing $x$ and $\xi$ the picture is very simple: there is only one independent GTMD. From eq.~\eqref{e:G_GTMD_S} it follows that in the limit of small $x$ and $\xi$ the GTMDs defined in eq.~\eqref{e:W_GTMD_par} are related as follows:
\begin{equation}
    \lim_{x,\xi \to 0} x\mathcal{F}_1 = \lim_{x,\xi \to 0} x\mathcal{F}_2^{(1)} = -4 \lim_{x,\xi \to 0} x\mathcal{F}_3^{(1)} = -2 \lim_{x,\xi \to 0} x\mathcal{F}_4^{(1)} = \mathcal{E}^{(1)} ,
    \label{e:relbetweenfcts}
\end{equation}
where we used the shorthand notation $\mathcal{F}_i^{(n)} \equiv [(\bm{k}^2 - \bm{\Delta}^2/4)/(2M^2)]^n \,\mathcal{F}_i$. We stress that the relation in eq.~\eqref{e:relbetweenfcts} only holds for the dipole-type gauge link structure and in the leading-twist picture we consider. In the forward limit (i.e.\ upon setting $\Delta = 0$), we recover the result for the (real) TMDs, $\lim_{x\to0} xf_1 = \lim_{x\to0} xh_1^{\perp(1)}$, which was already found in~\cite{Boer:2016xqr}. The Wilson loop GTMD $\mathcal{E}$ is related to the Wilson loop TMD $e$ defined in~\cite{Boer:2016xqr} as $\lim_{\Delta \to 0} \mathcal{E}(\bm{k}^2,\bm{\Delta}^2,\bm{k} \cd \bm{\Delta}) = e(\bm{k}^2)$.

The Wilson loop operator in eq.~\eqref{e:dipole_operator} can be written in terms of its real and imaginary parts:
\begin{equation}
    S^{[\Box]}(\bm{x},\bm{y}) = \mathcal{P}(\bm{x},\bm{y}) + i\mathcal{O}(\bm{x},\bm{y}) ,
    \label{e:S_real_im}
\end{equation}
where the pomeron and odderon operators are respectively given by
\begin{equation}
    \mathcal{P}(\bm{x},\bm{y}) \equiv \frac{1}{2N_c} \tr \left( U^{[\Box]} + U^{[\Box]\dag} \right) , \qquad \mathcal{O}(\bm{x},\bm{y}) \equiv \frac{1}{2iN_c} \tr \left( U^{[\Box]} - U^{[\Box]\dag} \right) .
    \label{e:P_and_O}
\end{equation}
The pomeron operator is C-even and T-even, whereas the odderon operator is both C-odd and T-odd. Following the notation of~\cite{Bomhof:2007xt}, the T-odd contribution to the dipole-type GTMD correlator in eq.~\eqref{e:G_GTMD_S} is given by 
\begin{align}
    G_{(d)}^{\text{(T-odd)}\,ij}(\bm{k},\bm{\Delta}) &\equiv \frac{1}{2} \left( G^{[+,-]\,ij}(\bm{k},\bm{\Delta}) - G^{[-,+]\,ij}(\bm{k},\bm{\Delta}) \right) \nn \\
    &= \frac{N_c}{\alpha_s} \left[ \frac{1}{2} \left( \bm{k}^2 - \frac{\bm{\Delta}^2}{4} \right) \delta_\st^{ij} + k_\st^{ij} - \frac{\Delta_\st^{ij}}{4} - \frac{k_\st^{[i} \Delta_\st^{j]}}{2} \right] \nn \\
    &\quad\, \times \left( G^{[\Box]}(\bm{k},\bm{\Delta}) - G^{[\Box^\dag]}(\bm{k},\bm{\Delta}) \right) \nn \\
    &= \frac{2iN_c}{\alpha_s} \left[ \frac{1}{2} \left( \bm{k}^2 - \frac{\bm{\Delta}^2}{4} \right) \delta_\st^{ij} + k_\st^{ij} - \frac{\Delta_\st^{ij}}{4} - \frac{k_\st^{[i} \Delta_\st^{j]}}{2} \right] \nn \\
    &\quad\, \times \int \frac{d^2\bm{x} \,d^2\bm{y}}{(2\pi)^4} \;e^{- i\bm{k} \cd (\bm{x} - \bm{y}) + i\bm{\Delta} \cd \frac{\bm{x}+\bm{y}}{2}} \left. \left\langle \mathcal{O}(\bm{x},\bm{y}) \right\rangle \right|_{\rm LF}.
    \label{e:odderon}
\end{align}
From the hermiticity and time reversal constraints, respectively given by
\begin{equation}
    G^{[\Box]*}(\bm{k},\bm{\Delta}) = G^{[\Box]}(\bm{k},-\bm{\Delta}) , \qquad G^{[\Box]*}(\bm{k},\bm{\Delta}) = G^{[\Box^\dag]}(-\bm{k},-\bm{\Delta}) ,
\end{equation}
it follows that the combination $G^{[\Box]}(\bm{k},\bm{\Delta}) - G^{[\Box^\dag]}(\bm{k},\bm{\Delta})$ appearing on the second line of eq.~\eqref{e:odderon} only contains odd powers of $\bm{k} \cd \bm{\Delta}$. Hence, the odderon contains only odd harmonics $\cos[(2n-1)(\phi_k - \phi_\Delta)]$, with $n \geq 1$. Note that in the forward limit, i.e.\ for $\Delta \to 0$, the T-odd contribution~\eqref{e:odderon} vanishes. This means that for unpolarized hadrons there are no odderon contributions in the TMD case; there is no spin-independent odderon for $\Delta = 0$. 

The Fourier transform of the GTMD correlator is the Wigner distribution
\begin{equation}
    W^{[\Box]}(\bm{b},\bm{k}) = \int d^2\bm{\Delta} \;e^{-i\bm{\Delta} \cdot \bm{b}} \;G^{[\Box]}(\bm{k},\bm{\Delta}) .
\end{equation}
The normalization of the correlator $G$ in eq.~\eqref{e:G_at_small_x} is $\int d^2\bm{k} \;G^{[\Box]}(\bm{k},\bm{\Delta}) = \delta^{(2)}(\bm{\Delta})$, or for the Wigner distribution $\int d^2\bm{k} \;W^{[\Box]}(\bm{b},\bm{k}) = 1$, independent of the impact parameter $\bm{b}$ of the nucleon. Indeed, for $\Delta = 0$ the nucleon expectation value in impact parameter space, $\left\langle S^{[\Box]} \left(\bm{b}+\tfrac{\bm{r}}{2},\bm{b}-\tfrac{\bm{r}}{2}\right) \right\rangle$ $\equiv$ $\bra{P} S^{[\Box]} \left( \bm{b}+\tfrac{\bm{r}}{2},\bm{b}-\tfrac{\bm{r}}{2} \right) \ket{P} / \braket{P|P}$, is independent of $\bm{b}$, as it would also be in an infinite nuclear environment with constant nucleon density. However, for a finite, sufficiently large, nucleus there is a dependence on $\bm{b}$, even if one takes $\Delta \approx 0$ for the nucleons inside the nucleus. Rather than a constant density one has a transverse impact parameter profile for $\bm{b}$ that reflects the nucleon density in a nucleus,
\begin{equation}
    T(\bm{b}) = \int_{-\infty}^\infty dz \;\rho_A \left( \sqrt{\bm{b}^2 + z^2} \right) ,
\label{e:profile}
\end{equation}
where $\rho_A(r)$ is the density of nucleons in the nucleus. For a heavy nucleus one typically takes a Woods-Saxon profile, which is slowly varying for $b$ values not too close to the edge of the nucleus. By assuming the $b$ dependence of the nuclear state to be approximately constant on the scale of the nucleon size, we effectively take $\ket{p'} \approx \ket{p} \approx \ket{P}$ for nucleons in the nucleus. In a large nucleus the Wigner distribution then becomes\footnote{In the next section, the gauge link dependence of $W$, $G$, and $S$ will be implicit, and we will suppress the subscript $A$ of $W$.}
\begin{equation}
    W_A^{[\Box]}(\bm{b},\bm{k}) = \int \frac{d^2\bm{r}}{(2\pi)^2} \;e^{-i\bm{k} \cdot \bm{r}} \left\langle S^{[\Box]} \left(\bm{b}+\tfrac{\bm{r}}{2},\bm{b}-\tfrac{\bm{r}}{2}\right) \right\rangle_{A},
\label{e:Wigner}
\end{equation}
where the nuclear averaged matrix element $\langle \ldots \rangle_A$ is evaluated between nucleon states at a given impact parameter $\bm{b}$ in the nucleus of which the distribution follows that of the normalized transverse impact parameter space profile function $T(\bm{b})$. In what follows, we will refer to $W_A$ in eq.~\eqref{e:Wigner} as the dipole Wigner distribution (and similarly for the dipole GTMD distribution). We already noted that in the forward limit $\Delta = 0$, where one integrates over all $\bm{b}$ values, odderon effects disappear. Similarly, in an infinite nucleus the $\bm{b}$ dependence becomes irrelevant and all odderon effects disappear. In a finite nucleus, however, double parton scattering effects in peripheral $pA$ collisions can retain odderon effects as will be discussed in the next section.

\section{Probing odd azimuthal correlations at small $\bm{x}$} \label{s:AzimuthalCorr.}
In this section we study odd azimuthal correlations in particle production in high-energy collisions originating from the dipole Wigner and GTMD distributions of gluons in the nucleus. We first consider the production of two hadrons in $pA$ collisions through double parton scattering. The source for azimuthal angular correlations (both even and odd) in the cross section of this process is the relative orientation between the transverse momentum $\bm{k}$ of the produced particle and its impact parameter $\bm{b}$ for an inhomogeneous nucleus, which is encoded in the dipole Wigner distributions. Secondly, we briefly comment on exclusive diffractive dijet production in ultra-peripheral $pA$ collisions. For this process the source of azimuthal correlations is given by the relative orientation between the transverse momentum $\bm{k}$ and the momentum transfer $\bm{\Delta}$, encoded in the dipole GTMDs. No odd-harmonic correlations in the cross section are identified in this case.

\subsection{Dihadron production through double parton scattering in $pA$ collisions}
We consider the production of two hadrons in $pA$ collisions at forward rapidity (in the fragmentation region of the proton) in the hybrid formalism~\cite{Dumitru:2005gt}. The produced hadrons probe the proton at large values of the longitudinal momentum fraction $x$ (the proton is dilute), while on the nucleus side the small-$x$ gluons govern the interaction (the nucleus is dense). The high gluon density state of the nucleus can be described by the CGC framework. We study double parton scattering with two incoming (valence) quarks from the proton that scatter off the CGC system, producing the two final-state hadrons. The distribution of the quarks from the proton is described by the double parton distribution (DPD) $F_p(x_1,x_2,\bm{b}_1-\bm{b}_2)$, with $x_1$ and $x_2$ denoting the longitudinal momentum fractions of the quarks and $\bm{b}_1$ and $\bm{b}_2$ denoting their scattering positions with respect to the center of the nucleus.

The transverse momentum of the final quarks is acquired through multiple rescatterings off the gluon field of the nucleus which are resummed into fundamental Wilson lines. The scattering of a quark at transverse position ${\bm{x}}$ in a covariant gauge and in the eikonal approximation is given by a fundamental Wilson line $U({\bm{x}}) \equiv U_{[-\infty,\infty;\bm{x}]}^n$, see eq.~\eqref{e:U^n}. The operator that describes the scattering of two quarks off the CGC at the level of the cross section is given by
\be 
    \left< \frac{1}{N_c} \tr \left[ U(\bm{x}_1) \,U^\dagger(\bm{y}_1) \right] \frac{1}{N_c} \tr \left[ U(\bm{x}_2) \,U^\dagger(\bm{y}_2) \right] \right>_{x,A} = \left< S(\bm{x}_1,\bm{y}_1) \,S(\bm{x}_2,\bm{y}_2) \right>_{x,A} ,
\ee
where $S$ is the dipole operator. The average $\left< \cdots \right>_{x,A}$ is the CGC average over the nucleus state, probed at some small, but finite, longitudinal momentum fraction $x$. Generalizing the single-inclusive cross section for quark production~\cite{Dumitru:2002qt}, one can write the cross section for the production of two quarks with momenta $\bm{k}_1$ and $\bm{k}_2$ as (see e.g.~\cite{Lappi:2015vta}):
\begin{align}
    \frac{d\sigma^{pA}}{dy_1 dy_2 \,d^2\bm{k}_1 d^2 \bm{k}_2} &\propto \int d^2\bm{b}_1 \,d^2 \bm{b}_2 \,F_p(x_1,x_2,\bm{b}_1-\bm{b}_2) \int \frac{d^2\bm{r}_1 d^2 \bm{r}_2}{(2\pi)^4} \;e^{- i\bm{k}_1 \cdot \bm{r}_1 - i\bm{k}_2 \cdot \bm{r}_2} \nn \\
    &\quad\, \times \left< S\left(\bm{b}_1+\tfrac{\bm{r}_1}{2},\bm{b}_1-\tfrac{\bm{r}_1}{2}\right) \,S\left(\bm{b}_2+\tfrac{\bm{r}_2}{2},\bm{b}_2-\tfrac{\bm{r}_2}{2}\right) \right>_{x,A} ,
    \label{e:crosssection_step1}
\end{align}
where $\bm{r}_1 \equiv \bm{x}_1-\bm{y}_1$ and $\bm{r}_2 \equiv \bm{x}_2-\bm{y}_2$ are the dipole orientations, and $\bm{b}_1 \equiv (\bm{x}_1+\bm{y}_1)/2$ and $\bm{b}_2 \equiv (\bm{x}_2+\bm{y}_2)/2$ are the corresponding impact parameters.\footnote{In our model we are neglecting quantum interference effects as discussed in~\cite{Kovner:2017vro,Kovner:2018vec}.} We will work in the large-$N_c$ limit and consider only the leading contributions in $1/N_c$ to the azimuthal correlations. At large $N_c$, the expectation value of the product of traces factorizes:
\begin{align}
    &\left< S\left(\bm{b}_1+\tfrac{\bm{r}_1}{2},\bm{b}_1-\tfrac{\bm{r}_1}{2}\right) \,S\left(\bm{b}_2+\tfrac{\bm{r}_2}{2},\bm{b}_2-\tfrac{\bm{r}_2}{2}\right) \right>_{x,A} \nn \\[5pt]
    \approx &\left< S\left(\bm{b}_1+\tfrac{\bm{r}_1}{2},\bm{b}_1-\tfrac{\bm{r}_1}{2}\right) \right>_{x,A} \left< S\left(\bm{b}_2+\tfrac{\bm{r}_2}{2},\bm{b}_2-\tfrac{\bm{r}_2}{2}\right) \right>_{x,A} .
    \label{e:approx_for_S^2}
\end{align}
The neglected corrections have been shown to give rise to azimuthal asymmetries in~\cite{Lappi:2015vta,Dusling:2017dqg,Dusling:2017aot}; here we demonstrate that odd azimuthal asymmetries arise already at leading power in $1/N_c$. Under this assumption, the cross section in eq.~\eqref{e:crosssection_step1} can been written as a convolution of two dipole Wigner distributions:
\begin{align}
    \frac{d\sigma^{pA}}{dy_1 dy_2 \,d^2\bm{k}_1 d^2 \bm{k}_2} &\propto \int d^2\bm{b}_1 \,d^2 \bm{b}_2 \,F_p(x_1,x_2,\bm{b}_1-\bm{b}_2) \int \frac{d^2\bm{r}_1 d^2 \bm{r}_2}{(2\pi)^4} \;e^{- i\bm{k}_1 \cdot \bm{r}_1 - i\bm{k}_2 \cdot \bm{r}_2} \nn \\
    &\quad\, \times \left< S\left(\bm{b}_1+\tfrac{\bm{r}_1}{2},\bm{b}_1-\tfrac{\bm{r}_1}{2}\right) \right>_{x,A} \left< S\left(\bm{b}_2+\tfrac{\bm{r}_2}{2},\bm{b}_2-\tfrac{\bm{r}_2}{2}\right) \right>_{x,A} \nn \\
    &= \int d^2\bm{b}_1 \,d^2 \bm{b}_2 \,F_p(x_1,x_2,\bm{b}_1-\bm{b}_2) \,xW(x,\bm{b}_1,\bm{k}_1) \,xW(x,\bm{b}_2,\bm{k}_2) ,
\label{eq:General_sigmaDPS}
\end{align}
where $W$ is defined in eq.~\eqref{e:Wigner}. 

To extract the angular correlations, we parametrize the Wigner distribution in terms of the different harmonic contributions~\cite{Hatta:2016dxp}:
\begin{align}
    xW(x,\bm{b},\bm{k}) &= x\mathcal{W}_0(x,{\bm{b}}^2,\bm{k}^2) + 2\cos(\phi_b-\phi_k) \,x\mathcal{W}_1(x,{\bm{b}}^2, \bm{k}^2) \nn \\
    &\quad + 2 \cos2(\phi_b-\phi_k) \,x\mathcal{W}_2(x,{\bm{b}}^2,\bm{k}^2) + \,\ldots
\label{eq:WignerParametrization}
\end{align}
The elliptic flow resulting from the elliptic Wigner distribution $x\mathcal{W}_2$ has been analyzed in~\cite{Hagiwara:2017ofm,Zhou:2016rnt,Iancu:2017fzn}. Here we focus on the odd-harmonic correlations generated by the imaginary part of the dipole scattering amplitude (the odderon). We will only consider the first odd contribution explicitly, i.e.\ the one associated to $x\mathcal{W}_1$.

We consider the angular asymmetries generated by the inhomogeneity of the nucleus in the transverse plane, which is naturally larger on the edge of the nucleus and hence our main focus is on peripheral collisions. For peripheral collisions involving an inhomogeneous nucleus, the distribution of quarks inside the proton in transverse space becomes important. For simplicity, we assume that the dependence on $\bm{b}_1-\bm{b}_2$ in the double quark distribution factorizes and we take a Gaussian ansatz for the transverse density profile~\cite{Blok:2010ge,Diehl:2011yj}:
\be 
    F_p(x_1,x_2,\bm{b}_1-\bm{b}_2) = f_p(x_1,x_2) \,\frac{1}{4\pi R_N^2} \;e^{-\frac{(\bm{b}_1-\bm{b}_2)^2}{4 R_N^2}} ,
\label{eq:Proton_DPD}
\ee
where $R_N$ denotes the radius of the proton. Using eqs.~\eqref{eq:WignerParametrization} and~\eqref{eq:Proton_DPD} in the cross section~\eqref{eq:General_sigmaDPS}, and integrating over the angles $\phi_{b_1}$ and $\phi_{b_2}$, we obtain
\begin{align}
    \frac{d\sigma^{pA}}{dy_1 dy_2 \,d^2\bm{k}_1 d^2 \bm{k}_2} &\propto \frac{\pi}{8R_N^2} \,f_p(x_1,x_2) \int db_1^2 \,db_2^2 \;e^{-\frac{b_1^2+b_2^2}{4R_N^2}} \nn \\
    &\quad\, \times \left[ 2 I_0\left(\frac{b_1b_2}{2R_N^2}\right) x\mathcal{W}_0(x,\bm{b}_1^2,\bm{k}_1^2) \,x\mathcal{W}_0(x,\bm{b}_2^2,\bm{k}_2^2) \right. \nn \\
    &\quad\, \left. + \;4\cos(\phi_{k_1}-\phi_{k_2}) \,I_1\left(\frac{b_1b_2}{2R_N^2}\right) x\mathcal{W}_1(x,\bm{b}_1^2,\bm{k}_1^2) \,x\mathcal{W}_1(x,\bm{b}_2^2,\bm{k}_2^2) \right. \nn \\
    &\quad\, \left. + \;4\cos2(\phi_{k_1}-\phi_{k_2}) \,I_2\left(\frac{b_1b_2}{2R_N^2}\right) x\mathcal{W}_2(x,\bm{b}_1^2,\bm{k}_1^2) \,x\mathcal{W}_2(x,\bm{b}_2^2,\bm{k}_2^2) \right] \nn \\
    &\quad\: + \,\ldots \,,
    \label{e:cross-section_final}
\end{align}
where $b_{1,2} \equiv |\bm{b}_{1,2}|$ and $I_n$ is the $n$th modified Bessel function of the first kind. We note that the odderon Wigner distribution $x\mathcal{W}_1$ of the nucleus gives rise to odd azimuthal correlations between the transverse momenta of the produced particles. The $\cos(\phi_{k_1}-\phi_{k_2})$ correlation comes with $x\mathcal{W}_1 \,x\mathcal{W}_1$, which is suppressed compared to the mixed terms of the form $x\mathcal{W}_0 \,x\mathcal{W}_1$ that however vanish upon integration over the impact parameter angles $\phi_{b_1}$ and $\phi_{b_2}$. One can also consider single-particle production and azimuthal correlations at fixed impact parameter between the direction of the produced particle in the transverse plane and the reaction plane. In the latter case one obtains nonvanishing odd angular coefficients from the mixed terms $x\mathcal{W}_0 \,x\mathcal{W}_1$. However, since $\bm{b}_1$ and $\bm{b}_2$ are not experimentally observable, we consider only two-particle correlations that survive the integration over the directions of the impact parameters.

The standard way of quantifying azimuthal correlations in particle production is through the flow coefficients $v_n$~\cite{Voloshin:1994mz}. For two-particle correlations, we have~\cite{Chatrchyan:2013nka}
\be 
    v_n(\bm{k},\bm{k}^\text{ref}) \equiv \frac{V_n(\bm{k},\bm{k}^\text{ref})}{\sqrt{V_n(\bm{k}^\text{ref},\bm{k}^\text{ref})}} ,
\label{e:vn}
\ee
where the coefficients $V_n$ are obtained from the decomposition of the cross section into Fourier modes in the relative azimuthal angle between the produced particles, and $\bm{k}^\text{ref}$ is a reference momentum corresponding to an experimental reference bin. The first odd coefficient, which characterizes the directed flow, is given by
\begin{equation}
    V_1(\bm{k}_1^2,\bm{k}_2^2) \equiv \frac{\int db_1^2 \,db_2^2 \;e^{-\frac{b_1^2+b_2^2}{4R_N^2}} \,I_1\left(\frac{b_1b_2}{2R_N^2}\right) x\mathcal{W}_1(x,\bm{b}_1^2,\bm{k}_1^2) \,x\mathcal{W}_1(x,\bm{b}_2^2, \bm{k}_2^2)}{ \int db_1^2 \,db_2^2 \;e^{-\frac{b_1^2+b_2^2}{4R_N^2}} \,I_0\left(\frac{b_1b_2}{2R_N^2}\right) x\mathcal{W}_0(x,\bm{b}_1^2,\bm{k}_1^2) \,x\mathcal{W}_0(x,\bm{b}_2^2,\bm{k}_2^2)} .
\label{eq:General_V1}
\end{equation}

We next derive an explicit expression for $V_1$ in the semi-classical picture of the CGC framework for small (but finite) $x$. We assume $x$ is not too small so that, as a first step, we can ignore quantum corrections from energy evolution. The expression for the angular-independent contribution, $x\mathcal{W}_0$, can be obtained from the real part of the dipole operator, $\mathcal{P}$ defined in~\eqref{e:P_and_O}, which has been calculated in the McLerran-Venugopalan (MV) model using a Gaussian distribution of color sources~\cite{Gelis:2001da}:
\begin{equation}
    \left< \mathcal{P}\left(\bm{b}+\tfrac{\bm{r}}{2},\bm{b}-\tfrac{\bm{r}}{2}\right) \right>_{x,A} = \exp\left[-\frac{1}{4} \,r^2 Q_s^2(b) \ln\frac{1}{r\Lambda}\right] ,
    \label{e:<P>}
\end{equation}
where the dipole size is $r \equiv |\bm{r}|$ and the scale $\Lambda$ serves as an infrared cutoff. The quark saturation scale $Q_s^2(b)$ characterizes the nonlinear gluon dynamics in the nucleus and can be defined in terms of the nuclear profile function $T(b)$ in eq.~\eqref{e:profile}:
\be 
    Q_s^2(b) \equiv \frac {4\pi \alpha_s^2 C_F}{N_c} \;T(b) .
\ee
For the nucleon number density $\rho_A(\vec{r})$ entering $T(b)$ one typically takes the Woods-Saxon distribution,
\begin{equation}
    \rho_A(r) = \frac{N_A}{1+\exp\left(\frac{r-R_A}{\delta}\right)} ,
\end{equation}
with $R_A = (1.12 \;\text{fm}) A^{1/3}$ the nuclear radius, $\delta = 0.54$ fm the width of the `nuclear edge'~\cite{Iancu:2017fzn}, and $N_A$ is a normalization factor such that $\int d^3\vec{r} \,\rho_A(\vec{r}) = A$, the number of nucleons inside the nucleus. For our numerical calculation of $v_1$ we will consider a lead nucleus with $A=208$ and a copper nucleus with $A=63$. To save computation time, we will use the leading asymptotic expansion for $I_i \left( \tfrac{b_1b_2}{2R_N^2} \right)$, since $b_1$ and $b_2$ both are taken larger than $b_0 \approx 1$ fm, and we will use
\be
    T(b) = \beta \left( 1 - \frac{1}{1 + e^{-(b/b_0 - \gamma)}} \right) ,
\ee
with $\beta= 2.13 \;\text{fm}^{-2}$ and $\gamma=5.4$ for $A=208$, and $\beta= 1.38 \;\text{fm}^{-2}$ and $\gamma=3.5$ for $A=63$, which approximates the Woods-Saxon distribution sufficiently well for our purpose.

Using eq.~\eqref{e:<P>}, the angular-independent Wigner function $x\mathcal{W}_0$ is given by
\begin{align}
    x\mathcal{W}_0(x,\bm{b}^2,\bm{k}^2) &= \int \frac{d^2\bm{r}}{(2\pi)^2} \;e^{-i \bm{k} \cdot \bm{r}} \left< \mathcal{P}\left(\bm{b}+\tfrac{\bm{r}}{2},\bm{b}-\tfrac{\bm{r}}{2}\right) \right>_{x,A} \nn \\
    &= \frac{1}{2\pi} \int_0^\infty dr \, r\, \exp\left[-\frac{1}{4} \,r^2 Q_s^2(b) \ln\frac{1}{r \Lambda}\right] J_0(k\,r) ,
\label{eq:W0}
\end{align}
where $k \equiv |\bm{k}|$ and $J_n$ is the $n$th Bessel function of the first kind.

\begin{figure}[!htb]
\centering
    \includegraphics[width=0.3\textwidth]{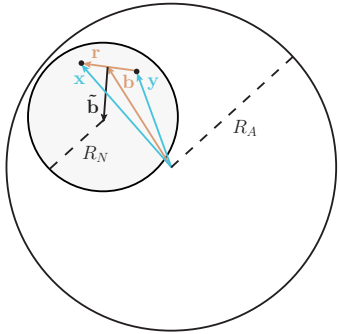}
\caption{Geometrical picture of the dipole hitting the nucleon inside the nucleus.}
\label{f:nucleus}
\end{figure}

Now we turn to the calculation of the first odd-harmonic Wigner function $x\mathcal{W}_1$ that can be obtained from the expectation value of the odderon operator $\mathcal{O}$ defined in~\eqref{e:P_and_O}. An analytical expression for the initial conditions of the impact parameter dependent odderon has been derived in~\cite{Kovchegov:2003dm,Hatta:2005as,Jeon:2005cf,Kovchegov:2012ga}. Here we employ the model result of~\cite{Kovchegov:2012ga} where the odderon was derived for a nucleus consisting of $A$ distinct nucleons. The expectation value of the odderon for a nuclear system is given by~\cite{Kovchegov:2012ga} 
\begin{align}
    \left< \mathcal{O}\left(\bm{b}+\tfrac{\bm{r}}{2},\bm{b}-\tfrac{\bm{r}}{2}\right) \right>_{x,A} &= - \frac{3}{4R_N^4} \,c_0\alpha_s^3
    r^2 \exp\left[-\frac{1}{4} \,r^2 Q_s^2(b) \ln\frac{1}{r\Lambda}\right] \nn \\
    &\quad\, \times \,\bm{r} \cdot \int d^2\tilde{\bm{b}} \;\tilde{\bm{b}} \;T\left(|\bm{b}+\tilde{\bm{b}}|\right) \theta(R_N-\tilde{b}) ,
\label{eq:O_General}
\end{align}
where $c_0 \equiv -(N_c^2-4)(N_c^2-1)/(12N_c^3)$. The impact parameter of the dipole with respect to the center of the struck nucleon is $-\tilde{\bm{b}}$ (and $\tilde{b} \equiv |\tilde{\bm{b}}|$), whereas $\bm{b}$ is the impact parameter with respect to the center of the nucleus, as before. The geometry of the scattering is illustrated in figure~\ref{f:nucleus}. The result in eq.~\eqref{eq:O_General} is expected to be a good approximation for $r < R_N$ (which means that the dipole is perturbatively small) and for $\tilde{b} \ll R_N$. 

We focus on peripheral collisions for which $\tilde{b} \ll b$, as $\tilde{b}$ is confined to the area of the nucleon. Hence, we can expand the nuclear profile function in powers of $\tilde{b}/b$:
\begin{equation}
    T\left(|\bm{b}+\tilde{\bm{b}}|\right) = \left[1+ \tilde{b}^i\frac{\partial}{\partial b^i} + \frac{1}{2} \,\tilde{b}^i \tilde{b}^j\frac{\partial^2}{\partial b^i \partial b^j} + \frac{1}{3!} \,\tilde{b}^i \tilde{b}^j\tilde{b}^k\frac{\partial^3}{\partial b^i \partial b^j \partial b^k} + \ldots \right] T(b) .
\label{eq:T_expansion}
\end{equation}
Plugging this expansion (up to second order in $\tilde{b}/b$) in eq.~\eqref{eq:O_General} and performing the integration over $\tilde{\bm{b}}$, we obtain:
\begin{align}
    \left< \mathcal{O}\left(\bm{b}+\tfrac{\bm{r}}{2},\bm{b}-\tfrac{\bm{r}}{2}\right) \right>_{x,A} &\approx -\frac{3\pi}{16} \,c_0 \alpha_s^3 r^2 \exp\left[-\frac{1}{4} \,r^2 Q_s^2(b) \ln\frac{1}{r \Lambda}\right] \frac{\bm{b} \cdot \bm{r}}{b} \;T'(b) \nonumber \\
    &= - \frac{3\pi}{16} \,c_0 \alpha_s^3 r^3 \exp\left[-\frac{1}{4} \,r^2 Q_s^2(b) \ln\frac{1}{r \Lambda}\right] \cos(\phi_b -\phi_r) \,T'(b) .
\end{align}
Note that only the linear term in the expansion survives the integration; it has given rise to a $\cos(\phi_b-\phi_r)$ correlation.\footnote{The quadratic term in the expansion gives rise to a $\cos2(\phi_b-\phi_r)$ correlation in the real part of the dipole operator, see~\cite{Iancu:2017fzn}.} From the Fourier transform of the odderon,
\begin{align}
    \int \frac{d^2\bm{r}}{(2\pi)^2} \;e^{-i \bm{k} \cdot \bm{r}} \left< \mathcal{O}\left(\bm{b}+\tfrac{\bm{r}}{2},\bm{b}-\tfrac{\bm{r}}{2}\right) \right>_{x,A} &= - \frac{3}{32} \,i c_0 \alpha_s^3 \cos(\phi_b-\phi_k) \,T'(b) \int_0^\infty dr \,r^4 \nn \\
    &\quad\, \times \exp\left[-\frac{1}{4} \,r^2 Q_s^2(b) \ln\frac{1}{r\Lambda}\right] J_1(k\,r) ,
\end{align}
one can extract $x\mathcal{W}_1$, as defined in eq.~\eqref{eq:WignerParametrization}:
\begin{equation}
    x\mathcal{W}_1(x,\bm{b}^2,\bm{k}^2) = \frac{3}{32} \,c_0 \alpha_s^3 \;T'(b) \int_0^\infty dr \,r^4 \,\exp\left[-\frac{1}{4} \,r^2 Q_s^2(b) \ln\frac{1}{r\Lambda}\right] J_1(k\,r) .
\label{eq:Wodd}
\end{equation}

\begin{figure}[!htb]
\centering
    \includegraphics[width=0.5\textwidth]{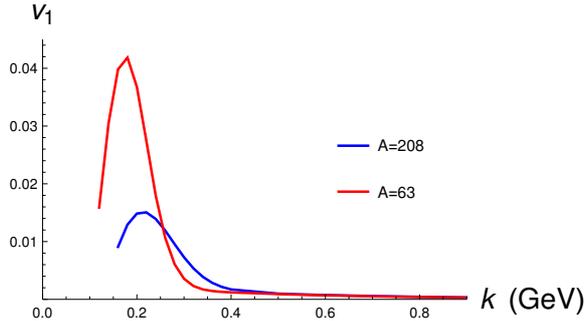}
\caption{The directed flow $v_1(k)$ for a lead nucleus with $A=208$ (blue curve) and for a copper nucleus with $A=63$ (red curve). We take $\alpha_s=0.3$, $\Lambda=0.24$ GeV, and $k^{\text{ref}}=0.8$ GeV.}
\label{f:v1}
\end{figure}

With the results in eqs.~\eqref{eq:W0} and~\eqref{eq:Wodd} we have obtained all the ingredients needed to calculate the directed flow $v_1$ defined in eqs.~\eqref{e:vn} and~\eqref{eq:General_V1} in the CGC framework. We perform an approximate numerical calculation to estimate the size of $v_1$ in our model. We calculate $v_1$ for a lead nucleus with $A=208$ and for a copper nucleus with $A=63$ in order to demonstrate the $A$ dependence of the result. Furthermore, we take $\alpha_s=0.3$ and $\Lambda = \Lambda_\text{QCD} = 0.24$ GeV. In figure~\ref{f:v1} we show $v_1$ as a function of $k$ for $k^{\text{ref}}=0.8$ GeV. Our result is strictly valid for perturbatively small dipoles, i.e.\ for $r<R_N \sim 1/\Lambda_\text{QCD}$, so we expect the calculation to break down for small momenta $k < \Lambda_\text{QCD}$. We plot $v_1$ starting from $k=0.12$ GeV for $A=63$ and $k=0.14$ GeV for $A=208$, as unphysical fluctuations are present at lower values of $k$ where it should go to zero as $k \to 0$. From the plot for a lead nucleus (blue curve) we infer that the observed sign change in the ATLAS data~\cite{Aad:2014lta} is not present in our result and we do not have the same behavior at large $k$.\footnote{Note that we do not include the sign function $\text{sgn}(k^{\text{ref}}-k^0)$ that is used in the definition of $v_1$ in~\cite{Aad:2014lta}. The sign function is defined to be negative for $k^{\text{ref}} < k^0=1.5$ GeV and is positive otherwise.} However, our goal here is not to describe the data, but rather to illustrate our main point with a simple model. It shows that the magnitude of $v_1$ for lower values of $k$ is of the same order as the maximal value observed in the data. We emphasize that we do not have any fit parameters in our result. The behavior and magnitude of $v_1$ are similar for other choices of $k^{\text{ref}}$. 

Figure~\ref{f:v1} shows that $v_1$ in our model calculation decreases with increasing $A$. This is expected, since the odderon contribution $xW_1$ has been computed with the MV model extended by a cubic action~\cite{Jeon:2005cf}, i.e. an action that includes a subleading correction in $1/A^{1/3}$ compared to the original MV model. Therefore, the whole observable is subleading in $1/A^{1/3}$, as expected for an effect that is proportional to the derivative of the nuclear thickness function $T(b)$. However, the precise dependence on $A$ is not straightforward to extract analytically from the explicit expressions because of the non-trivial $Q_s$ dependence.

Finally, let us comment that the odd harmonics computed in~\cite{Lappi:2015vta} in the nonlinear Gaussian approximation are suppressed in $1/N_c$ compared to our result, since we are considering the leading-order terms in the $1/N_c$ expansion in eq.~\eqref{e:approx_for_S^2} which do not produce odd harmonics in the mechanism considered in~\cite{Lappi:2015vta}.


The results so far have been derived for the production of two hadrons by two incoming quarks from the proton that scatter off the nuclear CGC. This description is valid at forward rapidity, such that the proton consists of valence quarks only. Moving away from the forward-rapidity region, one needs to account for the presence of sea quarks and antiquarks in the proton's wave function. The scattering of an antiquark is given by the complex conjugate of the quark's dipole operator, which implies that $x\mathcal{W}_1^\text{quark} = -x\mathcal{W}_1^\text{antiquark}$. Hence, if one assumes that away from the valence region the proton consists of an equal number of quarks and antiquarks with exactly the same DPDs for $qq$, $\bar{q}\bar{q}$, and $q\bar{q}$ pairs, the sum over all partons scattering off the nucleus would not give any odd harmonics. However, it is known that for values of $x \leq 10^{-1}$, the DPDs differ for different pairs of partons with an $x$-dependent width in the Gaussian model~\cite{Diehl:2004cx,Diehl:2014vaa}. In general, the proton DPD takes the following form:
\be 
    F_p(x_1,x_2,\bm{b}_1-\bm{b}_2) = f_{ab}(x_1,x_2)\, \frac{1}{4\pi h_{ab}(x_1,x_2)} \;e^{-\frac{(\bm{b}_1-\bm{b}_2)^2}{4 h_{ab}(x_1,x_2)}} ,
\label{eq:Proton_DPD_xDependent}
\ee
where the labels $a,b$ refer to quarks or antiquarks. The precise form of $h_{ab}$, including numerical values for the relevant parameters, can be found in~\cite{Diehl:2014vaa}. Due to the fact that the proton DPDs differ for different combinations of quarks and antiquarks, our mechanism would \emph{also} produce nonzero odd azimuthal correlations away from the forward-rapidity region. For the case of gluon production, the odd harmonics in our model are zero by construction. This is because the adjoint dipole operator $S_A = \tr [V(\bm{x}) \,V^\dagger(\bm{y})] /(N_c^2-1)$, describing the scattering of gluons off the CGC, is purely real. 

\subsection{Diffractive dijet production in ultra-peripheral $pA$ collisions} 
The gluon dipole GTMD distribution (and therefore the dipole Wigner distribution) appears also in the cross section for diffractive dijet production in DIS and in ultra-peripheral $pA$ collisions, which has been shown in~\cite{Altinoluk:2018hcu,Hatta:2016dxp,Hagiwara:2017fye}. However, the mechanism we have discussed so far does not generate odd azimuthal angular correlations between the produced jets. The cross section for this process involves the absolute value squared of the dipole operator and the odderon only gives corrections to the angle-independent and $\cos 2(\phi_R-\phi_\Delta)$ coefficients. For completeness we derive the form of these corrections.

Analogous to the parametrization in eq.~\eqref{eq:WignerParametrization} in terms of Wigner distributions, the dipole GTMD correlator, defined in eq.~\eqref{e:G_at_small_x}, can be parametrized as
\begin{align}
    G(x,\bm{k},\bm{\Delta}) &= \mathcal{G}_0(x,\bm{k}^2,\bm{\Delta}^2) + 2\, i \cos(\phi_k-\phi_\Delta) \,\mathcal{G}_1(x,\bm{k}^2,\bm{\Delta}^2) \nn \\
    &\quad\, + 2 \cos2(\phi_k-\phi_\Delta) \,\mathcal{G}_2(x,\bm{k}^2,\bm{\Delta}^2) + \,\ldots
\label{eq:GTMDParametrization}
\end{align}
The produced jets have transverse momenta $\bm{k}_1$ and $\bm{k}_2$ such that $\bm{k}_1 + \bm{k}_2 = -\bm{\Delta}$, and the relative transverse momentum of the dijet is given by $\bm{R} \equiv (\bm{k}_2-\bm{k}_1)/2$. The contributions from the functions $\mathcal{G}_0$ and $\mathcal{G}_2$ to the dijet production cross section have been calculated in~\cite{Hagiwara:2017fye}:
\be
    \frac{d\sigma^{pA}}{dy_1 dy_2 \,d^2\bm{k}_1 d^2\bm{k}_2} \propto A^2 + 2 \cos 2(\phi_R-\phi_\Delta) \,AB ,
\label{e:cross-section_withoutG1}
\ee
where 
\begin{align}
    A(x,\bm{R}^2,\bm{\Delta}^2) &\equiv -\int_0^R dk \, k \;\mathcal{G}_0(x,\bm{k}^2,\bm{\Delta}^2) , \\
    B(x,\bm{R}^2,\bm{\Delta}^2) &\equiv - \int_0^R dk \,\frac{k^3}{R^2} \;\mathcal{G}_2(x, \bm{k}^2,\bm{\Delta}^2) + \int_R^\infty dk \, \frac{R^2}{k} \;\mathcal{G}_2(x, \bm{k}^2,\bm{\Delta}^2) ,
\end{align}
with $k \equiv |\bm{k}|$ and $R \equiv |\bm{R}|$. The odd function $\mathcal{G}_1$ modifies the cross section in eq.~\eqref{e:cross-section_withoutG1} to
\be
    \frac{d\sigma^{pA}}{dy_1 dy_2 \,d^2\bm{k}_1 d^2\bm{k}_2} \propto \left( A^2 + \frac{1}{2} \,C^2 \right) + 2 \cos 2(\phi_R-\phi_\Delta) \left(AB +\frac{1}{4} \,C^2 \right) ,
\label{e:cross-section_withG1}
\ee
where
\be 
    C(x,\bm{R}^2,\bm{\Delta}^2) \equiv - \int_0^R dk \,\frac{k^2}{R} \;\mathcal{G}_1(x,\bm{k}^2,\bm{\Delta}^2) + \int_R^\infty dk \,R \;\mathcal{G}_1(x, \bm{k}^2,\bm{\Delta}^2) .
    \label{e:C_coefficient}
\ee
We infer from eq.~\eqref{e:cross-section_withG1} that the GTMD odderon $\mathcal{G}_1$ only corrects the size of the even harmonics and does not lead to odd harmonics.

\section{Discussion and conclusions} \label{Conclusions}
We have provided an alternative parametrization of the gluon-gluon GTMD correlator for unpolarized hadrons in terms of definite-rank GTMDs of leading twist. For the dipole-type gauge link structure and for vanishing $x$ and $\xi$ this correlator is related to the correlator of a Wilson loop. The fact that the Wilson loop correlator can be parametrized in terms of just a single GTMD, implies that the dipole GTMDs all become proportional to each other in this particular limit. Hence, one expects that in the small-$x$ kinematic region where gluon effects dominate those of quarks, the picture of gluon GTMDs becomes very simple. The gluon GTMDs with a dipole-type gauge link contain both C-even (pomeron) and C-odd (odderon) contributions, the latter disappearing in the forward limit where GTMDs reduce to TMDs.

The odderon contribution to the gluon GTMD (in a situation without polarization) originates from correlations between the off-forwardness and the gluon momentum in the transverse plane. In practice this means that we need to look for correlations between the impact parameter $\bm{b}$ of nucleons and the noncollinearity of gluon operators appearing in the transverse size $\bm{r}$ of Wilson lines, in particular the Wilson loop. Odderon effects will then show up as odd flow coefficients in $pA$ collisions. Using a simple model for dihadron production from double parton scattering in peripheral $pA$ collisions, we have shown that the odderon Wigner distribution gives rise to odd azimuthal asymmetries, even in the large-$N_c$ limit and for a radially symmetric nucleus. At leading order in $1/N_c$, the cross section of this process is proportional to a convolution of two dipole Wigner distributions which encode partonic correlations within the nuclear wave function. Those distributions give rise to both even and odd harmonics in the two-particle angular correlations, the latter being a double C-odd effect. Within the CGC model, we have calculated the directed flow coefficient $v_1$ that quantifies the contribution of the first harmonic with respect to the angular independent contribution. For peripheral collisions involving a lead nucleus described by a Woods-Saxon-like profile, we find a $v_1$ at the percent level. This order of magnitude is consistent with the ATLAS measurements~\cite{Aad:2014lta}, although the shape of $v_1$ is not well-described using the simple model. We point out that a nonzero $v_3$ can be generated from higher-order corrections to eq.~\eqref{eq:O_General} and from the third derivative of the profile function in eq.~\eqref{eq:T_expansion}.

It would be interesting to see how quantum effects from small-$x$ evolution modify our result derived in the classical approach. In the CGC framework, one can calculate the small-$x$ evolution of $x\mathcal{W}_0$ and $x\mathcal{W}_1$ from the nonlinear Jalilian-Marian-Iancu-McLerran-Weigert-Leonidov-Kovner (JIMWLK) equation~\cite{JalilianMarian:1997jx,JalilianMarian:1997gr,JalilianMarian:1997dw,Kovner:1999bj,Kovner:2000pt,Weigert:2000gi} applied to the real ($\mathcal{P}$) and imaginary ($\mathcal{O}$) parts of the dipole operator, respectively. The small-$x$ evolution of $v_3$ has been considered in~\cite{Lappi:2015vha,Lappi:2015vta} and they observed that $v_3$ decreases with decreasing $x$, while~\cite{Dumitru:2014vka} found that $v_1$ and $v_3$ increase with decreasing $x$ for small values of the dipole size. We leave the analysis of quantum corrections to the odd angular coefficients for the future.

\acknowledgments
We thank Sabrina Cotogno and Jonathan Gaunt for useful discussions and Miguel Garc\'ia Echevarr\'ia for his contributions to the initial stage of this project, in particular the derivation of eq.~\eqref{e:C_coefficient}. This research is in part supported by the European Community under the ``Ideas'' program QWORK with contract no. 320389.

\bibliographystyle{JHEP}
\bibliography{references}

\end{document}